%% file: ms.tex
\documentclass[twocolumn]{emulateapj}
\usepackage{graphicx}
\usepackage{epsfig}
\usepackage{amsmath}
\usepackage{natbib}
\usepackage[dvips]{color}
\newcommand{\msun}{M_{\odot}}
\newcommand{\acrvs}{$\cal{A}$$_{CRV}~$}
\newcommand{\acrv}{$\cal{A}$$_{CRV}$}

\begin{document}
\title{Kinematic Signatures of Subvirial Initial Conditions in Young Clusters}
\author{Eva-Marie Proszkow}
\affil{Michigan Center for Theoretical Physics, Physics Department, University of Michigan, Ann Arbor, MI 48109;\\ emdavid@umich.edu}
\author{Fred C. Adams}
\affil{Michigan Center for Theoretical Physics, Physics Department, University of Michigan, Ann Arbor, MI 48109; \\Astronomy Department, University of Michigan, Ann Arbor, MI 48109}
\author{Lee W. Hartmann}
\affil{Astronomy Department, University of Michigan, Ann Arbor, MI 48109}
\and
\author{John J. Tobin}
\affil{Astronomy Department, University of Michigan, Ann Arbor, MI 48109}

\shorttitle{Kinematic Signatures}
\shortauthors{Proszkow et al.}

\begin{abstract}
Motivated by kinematic observations of young embedded clusters, this paper explores possible kinematic signatures produced by asphericity and departures from initial virial equilibrium in these systems.  Specifically, the kinematic quantity that is measured and calculated in this study is the distribution of the line-of-sight velocities as a function of position along the cluster. Although clusters are found within a wide range of sizes, we focus on the regime with stellar membership $N \sim 10^3$. The gravitational potential of these young clusters is dominated by the gas, and the geometry of the gas distribution is generalized to include axisymmetric (and triaxial) forms. With this loss of symmetry, the kinematic results thus depend on viewing angle. This work also considers the stars to begin their trajectories with subvirial speeds, as indicated by observations of core motions in such clusters. Our results determine the conditions necessary for the kinematic signature to display interesting structure, i.e., a non-spherical potential, a viewing angle that is not along one of the principal axes, and subvirial starting conditions.  We characterize the effects on this signature due to projection angle, initial stellar velocities, cluster elongation, and star formation efficiency. Finally, we compare our theoretical results to recent kinematic observations of the Orion Nebula Cluster; we find that the observations can be explained provided that the cluster is non-spherical, starts with subvirial initial velocities, and is not viewed along a principal axis. 
\end{abstract}

\keywords{open clusters and associations: general, stellar dynamics, methods: n-body simulations, stars: formation}

\section{Introduction} \label{sec:Intro}

Most stars are thought to form within clusters of some type \citep[e.g.,][]{Lada2003ARAA,Porras2003AJ}, although a great deal of controversy remains concerning the distribution of cluster properties.  Nonetheless, given that stars form in clusters, two overarching questions immediately arise. The first considers the clusters as astrophysical objects: [1] How can molecular clouds produce aggregates of $N > 100$ stars with centrally concentrated surface density, with the massive stars near the center, and with a stellar mass distribution that follows the IMF, all within a few pc and within a few Myr? A second vital question then becomes: [2] If stars form in clusters, how does the cluster environment affect star formation and the accompanying process of planet formation? A complete understanding of star and planet formation requires detailed answers to both questions. However, this paper will focus on one specific issue within this larger context. Observed young embedded stellar clusters display departures from both spherical symmetry and initial virial equilibrium, and this paper will explore the effects of these complications on the kinematics of cluster members through $N$-body simulations. These results, in turn, will help provide a contribution to the overarching questions posed above.

Departures from spherical symmetry in star forming regions have long been known, but have been little studied. On the scale of cluster themselves (with radii of a few pc, and masses of 100 -- 3000 $M_\odot$), for example, some systems are categorized as irregular \citep{Lada2003ARAA}, or filamentary \citep{Walsh2004ApJ}, or otherwise described as non-spherical \citep{Allen2007PPV}. Furthermore, the star formation efficiencies are low, so that the gravitational potential in these systems is dominated by the mass of the gas (at least in the early phases of evolution, before gas expulsion). As a result, the gas potential must be generalized to include the observed departures from spherical symmetry (see \citealt{Adams2007ApJ} and \S 2). For completeness, we also note that on the smaller scale of cores (or clumps or kernels) that represent individual star formation events, the pre-collapse gas is observed to be non-spherical, with typical aspect ratios of 2:1 \citep[e.g.,][]{Myers1991ApJ,Ryden1996ApJ}. 

Observations are starting to show that clusters begin their evolution with subvirial initial conditions, and hence this complication must be included in our simulated clusters. In many regions, pre-stellar clumps are observed to move subsonically before the clumps form stars \citep{Peretto2006AA,Walsh2004ApJ,Andre2002ApSS,Kirk2007ApJ}, implying that newly formed stars begin their dynamical evolution with subvirial speeds. Motivated by such observations, our work considers clusters that are seeded with subvirial stellar velocities.  As such, our  N-body simulations (see \S 2) differ from those of many preceding studies \citep[though not all, see][]{Bonnell1998MNRAS,Adams2006ApJ} that assume the initial phase space variables of the stars are close to virial equilibrium.   Subvirial initial conditions can have a significant impact on the early cluster evolution \citep{Adams2006ApJ}, and are thus considered herein.

The theoretical work presented here provides a determination of the kinematic signature of young embedded clusters with both subvirial starting conditions and non-spherical gas potentials. Fortunately, astronomical observations are now becoming available to compare with these theoretical calculations. In this paper, we focus on clusters of moderately large size, with stellar membership $N \sim 10^3$, appropriate for the Orion Nebula Cluster (ONC). Clusters of this size are large enough for interesting kinematic signatures to arise in observations, and small enough that many such clusters are expected within the galactic cluster population. In addition, the ONC is an example of a nonspherical cluster as the stellar population is elongated North to South along the filament of dense molecular gas in the region \citep{Hillenbrand1998ApJ}.  Recent observations in the ONC \citep[Tobin et al. 2009, submitted to ApJ, TO9 hereafter]{Furesz2008ApJ} display a gradient in the radial velocity structure along the length of the cluster, much like that expected for the elongated subvirial clusters considered herein.  

This paper is organized as follows. In \S 2, we outline the theoretical approach used in this paper. Specifically, we describe the $N$-body codes, the required number of realizations of the numerical experiments, the implementation of subvirial starting speeds, the inclusion axisymmetric and triaxial gas distributions, as well as the specification of simulation parameters.  We then present the theoretical results of our simulations in \S 3, with a focus on the velocity signatures produced by the departures from equilibrium and spherical symmetry. In \S 4, we compare our results to observations, primarily the kinematic velocity measurements recently carried out in the ONC. We find good qualitative agreement, and reasonable quantitative agreement, and suggest that the observed kinematic signature requires subvirial starting conditions, non-spherical potentials, and viewing angles that do not coincide with the principal axes of the systems. These conclusions and other results are summarized in \S 5, along with a discussion of their implications.

\section{Methods} \label{sec:Methods}

$N$-body simulation techniques are used to calculate the dynamics of young stellar clusters during the embedded star forming epoch.  We consider clusters with nonspherical geometries and subvirial initial velocities and study the observed kinematics of these clusters.  In this section, we discuss the $N$-body code used, the simulation parameters chosen, and the experiments performed to identify and characterize kinematic signature in nonspherical clusters with subvirial initial conditions.

	\subsection{N-Body Simulation Techniques}
		
    The dynamical evolution of a young embedded cluster depends on its initial stellar distribution, the distribution and removal mechanism of the embedding gas, the star formation history, and is especially sensitive to the initial velocity distribution \citep[e.g.,][]{Adams2006ApJ}.  In this paper we complete a suite of $N$-body simulations of young embedded clusters to understand how the cluster's initial spatial and velocity distribution imprints itself on the evolved cluster's kinematic structure. The NBODY2 direct integration code developed by \cite{Aarseth2001NewAstro} is used as a starting point to calculate the cluster's dynamics from the star-forming epoch out to ages of $10$ Myr.  As outlined below, we modify the code to include specific stellar and gas distributions and star formation epochs that are similar to those observed in young embedded clusters. 
    
    Distributions of stellar positions and velocities are required to characterize kinematic signatures in these young clusters. We find that $\sim 50$ equivalent realizations of each set of cluster initial conditions are required to provide statistically robust distributions of stellar positions and radial velocities. This study focuses on the kinematic signatures produced by two different contributions to the initial conditions commonly present in young embedded clusters: asphericity (combined with effects due to projection of a 3D cluster onto the 2D sky), and subvirial initial velocities.  In addition, we consider the effect of extinction on the observed kinematic structure of young clusters.  
    
    Cluster evolution is numerically integrated from the star formation epoch ($0-1$ Myr) through the gas removal phase (at $t = 5$ Myr) out to ages of $10$ Myr (see \S \ref{sec:SimParams} for further detail).  After $50$ realizations of each set of initial conditions are completed, the results are combined to produce distributions of stellar positions and velocities.  The simulations provide three dimensional position and velocity information for each stellar member at intervals of $0.1$ Myr.  These six phase space coordinates are reduced to two position coordinates in the plane of the sky and two velocity components, one along the line of sight (radial velocity) and one in the plane of the sky (transverse velocity).  For the sake of definiteness, the $\hat{z}'$ projected axis is defined by the projection of the three dimensional $\hat{z}$ axis (the major axis in elongated clusters) onto the plane perpendicular to the line of sight. The terms ``North'' and ``South'' with reference to projections of simulated clusters correspond to positive and negative $\hat{z}'$ values respectively.  This nomenclature is arbitrary and has been chosen to coincide with that of the ONC, which is elongated North to South and displays North-South asymmetry in the radial velocity structure (see \S \ref{sec:CompareObs}).

	\subsection{Simulation Parameters}\label{sec:SimParams}
		
	\emph{Cluster Membership $N$:} We consider moderately large young clusters comparable in size to the Orion star forming region.  Specifically, our simulated clusters contain $N = 2000$ or $2700$ stars.  Estimates of the stellar population  of the ONC vary depending on the cluster radius adopted, and our choices of $N$ roughly reflect the range of this variation (e.g. \cite{Hillenbrand1998ApJ} advocate N = 2300, near the center of this range). We found that there were no significant differences in the kinematic signature observed in clusters of $2000$ and $2700$ stars.  Another motivation for our choice of cluster size is that clusters of this size have large enough memberships that kinematic signatures may be identified in the data from observed clusters and compared to the results presented in this study.
	
	The stellar masses are sampled from the log-normal analytic fit presented by \cite{Adams1996ApJ} to the standard IMF of \cite{Miller1979ApJS}.  The masses range from $0.07$ to $10 \msun$.  A limited number of simulations with equal stellar masses were completed and we found that the kinematic signature discussed herein was present in both the single stellar mass clusters and those with a more realistic stellar mass distribution.
		
	\emph{Stellar Distribution:} One goal of this study is to compare kinematical signatures of clusters with spherical geometries to those of elongated clusters.  Spherical clusters have centrally concentrated stellar distributions described by a $r^{-1}$ density profile.  Two elongated stellar distributions are considered: uniform density prolate spheroids and centrally concentrated clusters with  $\rho \sim m^{-1}$, where $m^2 = x^2/a^2+y^2/b^2+z^2/c^2$ is the triaxial coordinate. We compare elongated clusters with different aspect ratios to characterize the effect of cluster elongation on the observed kinematic signature.  Table \ref{tab:StarDist} summarizes the initial stellar density distributions used in the cluster simulations.
	
	\emph{Initial Speeds:} Initial stellar velocities are sampled from a uniform distribution within a unit sphere, producing an isotropic and position-independent velocity distribution.  The velocities are then scaled to produce a cluster in a particular virial state, defined by the virial ratio $Q \equiv |K/W|$ where $K$ is the total kinetic energy of the cluster and $W$ is the total virial potential energy of the cluster. A cluster in virial equilibrium is defined to have $Q = 0.5$.  Simulations of clusters with virialized initial velocities are compared to clusters with subvirial initial velocities. The subvirial clusters are chosen to have $Q = 0.04$ or $0.15$, which corresponds to initial velocities that are approximately one third or one half of the virial velocity, respectively.
	
	\emph{Star Formation History:} The simulated clusters have a star formation epoch that lasts for the first $1$ Myr of the cluster's evolution.  During simulation initialization each star, regardless of mass or initial position within the cluster, is assigned a random formation time between $0$ and $1$ Myr (with a uniform distribution, independent of stellar mass, over this interval).  A star is then tied to its formation site (chosen as described above) until its collapse phase of star formation is complete. That is, the stars are initially included in the simulation as point masses, but are held at a fixed position until their formation time. After that time, the star becomes free to move through the total gravitational potential of the cluster with an initial velocity sampled from the distribution described above. As a result, the stars do not execute ballistic orbits until after they have formed. For completeness, we note that stellar evolution is not included in these simulations.
	
	We vary the star formation efficiency (SFE) from $17\%$ to $50\%$, which spans the range of mass estimates of the gas in the region of the ONC.  It is important to note that although the evolution of a cluster subsequent to gas removal depends strongly on the SFE, its evolution prior to gas removal is more sensitive to the virial ratio than it is to the SFE alone.  In other words, provided that the stars are moving with sufficient velocities to account for the additional potential due to excess gas (i.e., their virial ratios are comparable), the evolution of a cluster with an SFE of $33\%$ or $50\%$ will be qualitatively similar during the embedded stage.  Increasing the SFE will, however, decrease the average virial velocity.  After gas is removed from the cluster, the subsequent dynamical evolution does depend on the SFE. Of course, increasing the SFE will increase the rate of close encounters, but this effect is modest on the (short) time scales of interest here \citep{Adams2006ApJ}.
	
	\emph{Embedding Gas Distributions:}  We assume the distributions of stars in a cluster roughly traces the geometry and density of the embedding gas.  Thus a spherical centrally concentrated Hernquist profile \citep{Hernquist1990ApJ} is chosen to represent the embedding gas in a spherical cluster \citep[see][]{Adams2006ApJ}.  
	
	Likewise, the elongated stellar distributions are embedded in elongated gas potentials.  Specifically, the uniform density stellar distributions are embedded in a uniform density gas distribution that is twice the extent of the stellar distribution.  The homoeoid theorem states that the net force on a particle within a uniform density homoeoid shell is zero. Thus the larger gas distribution allows for a simpler computation of the force and potential terms due to the embedding gas without changing the dynamics of the system \citep{BT1987Book}.
	
	The centrally concentrated prolate clusters are embedded in a static gas potential of the form $\rho \sim m^{-1}$, where $m$ is the generalized coordinate.  Calculation of the force terms, analytic expressions for the potential, and a discussion of orbits and orbit instabilities within this triaxial potential were presented in \cite{Adams2007ApJ}. The gas distributions and associated parameters are summarized in Table \ref{tab:GasDist}. 

  Observations indicate that embedding gas does not remain in young clusters for a long time.  After a few million years, winds from hot young stars begin to carve out the embedding gas and very few embedded clusters are found with ages greater than $\sim 5$ Myr.  Our simulations account for gas removal as a temporal step function in the evolution of the static gas potential: at $t = 5$ Myr, the gas potential is completely removed from the cluster which then continues to evolve due to interactions between the stars.  After gas removal, the cluster expands and a significant fraction of the members become gravitationally unbound.

\input{tab1.tex}

\input{tab2.tex}

  \subsection{Numerical Experiments}
	 In this section we discuss the specific parameters that were varied to study the effects of particular initial conditions on the kinematic structure observed in the clusters.  A more detailed discussion of each experiment's initial set-up and result is reserved until \S\ref{sec:SimResults}.
	 
	 \emph{Cluster Geometry and Virial Ratio:} We compare the evolution of clusters under various assumptions of initial geometry and virial balance.  Specifically, we compare centrally concentrated ($\rho \sim r^{-1}$) spherical clusters with subvirial ($Q = 0.04$) and virial ($Q = 0.5$) initial velocities to uniform density elongated clusters with similarly subvirial and virial velocities.  The spherical clusters are $2$ pc in radius, and the elongated clusters have axis parameters $a = b = 2$ pc and $c = 4$ pc. Each of these clusters had stellar membership $N = 2000$ and an SFE of $50\%$.  We find that only clusters with subvirial initial velocities \emph{and} elongated geometries produce significant gradients in the radial velocity along the length of the cluster.  The observed radial velocity gradients are thus a combined effect of global collapse and the projection of a non-spherical cluster.  Hereafter, the term `kinematic signature' refers to this radial velocity gradient. This kinematic signature is discussed in detail in \S \ref{sec:RVStructure}.
	 		
	With the requirements of subvirial velocities and elongated stellar/gas distributions identified as prerequisites for the kinematic signature, we proceed to complete a series of numerical experiments to determine how changing cluster parameters in subvirial elongated clusters changes the observed structure of the signature.
		  
	\emph{Initial Virial Ratio:}  To characterize the effect of subvirial velocities on the kinematic signature, the evolution of centrally concentrated elongated clusters with virial ratio $Q$ ranging from $0.04$ to $0.15$ are compared.  These virial ratios correspond to average initial velocities that are roughly one third to one half of the virial velocity and are comparable to pre-stellar clump velocities observed many star forming regions including NGC 2264 \citep{Peretto2006AA} and Perseus \citep{Kirk2007ApJ}.
		
	\emph{Cluster Elongation:}  The effect of cluster elongation on the evolved velocity structure of the cluster is studied by comparing subvirial clusters that range from spherical to elongated with aspect ratios ranging from $1$ to $4$. Recent results from the \emph{Spitzer} Young Cluster Survey indicate this range of aspect ratios is appropriate, as the clusters in the survey had aspect ratios between $0.53$ and $3.88$ (Gutermuth et al. 2008, in preparation).
			
	\emph{Initial Density Distribution:}  Another experiment compares subvirial elongated clusters with uniform density distributions to those with $\rho \sim m^{-1}$ to study the effect of the density distribution on the strength of and evolution of the kinematic signature.
		 
	\emph{Star Formation Efficiency:}  To determine the effect of SFE on the kinematic signatures observed in the cluster, we compare subvirial elongated clusters with efficiencies ranging from $17$ to $50\%$.

\section{Simulation Results} \label{sec:SimResults}

	\subsection{Radial Velocity Structure Due to Global Collapse and Elongation}\label{sec:RVStructure}

	We find that radial velocity gradients along the length of the clusters are produced by a combination of two effects: [1] projection of an elongated or non-spherical cluster and [2] subvirial initial velocities.  The separation of these two effects is nontrivial, as discussed below.
		
\begin{figure}
  \epsscale{1.0}
  \plotone{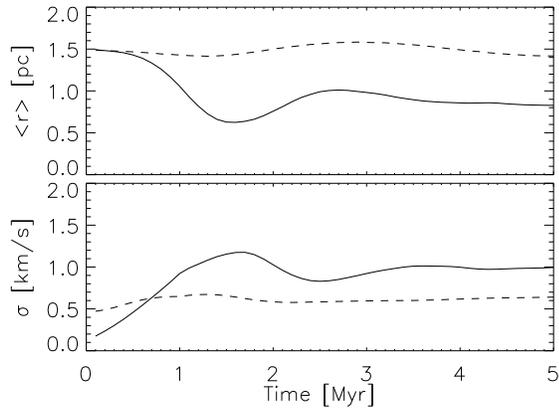}
	\caption{The top panel displays the evolution of the mean radius of a spherical cluster with virial (dashed curve) and subvirial (solid curve) initial velocities. The bottom panel displays the evolution of the velocity dispersion of a spherical cluster with virial (dashed curve) and subvirial (solid curve) initial velocities.}
	\label{fig:DispvsTime}
\end{figure}

  A cluster that is initially subvirial will collapse as stars in the outer parts of the cluster fall toward the cluster's center of mass.  This collapse takes place because cluster members that are seeded with subvirial velocities are not moving fast enough to remain in orbit (at their starting radial positions) around the cluster's center.  Instead they fall through the gravitational potential, gain kinetic and potential energy, and eventually reach an equilibrium state in which the virial theorem is satisfied. Thus, during the first crossing time, a subvirial cluster will collapse significantly as the stellar velocities increase. Figure \ref{fig:DispvsTime} compares the mean cluster radius and velocity dispersion as a function of time for spherical clusters with both virial and subvirial initial conditions.  The overall cluster collapse and velocity enhancement in subvirial clusters is clearly demonstrated by these plots.
  
  During the initial collapse phase, the velocity vectors of the cluster members are preferentially directed toward the center of mass until the stars pass close to the cluster's center and continue on (mostly) radial trajectories outwards.  During this first crossing time, some close encounters may cause the individual stellar motions to deviate from this general pattern.  As shown in \cite{Adams2006ApJ}, however, close encounters are relatively rare on these short time scales (a few Myr or less) and the aggregate dynamics are thus dominated by this initial collapse and re-expansion.

\begin{figure}
  \epsscale{1.0}
  \plotone{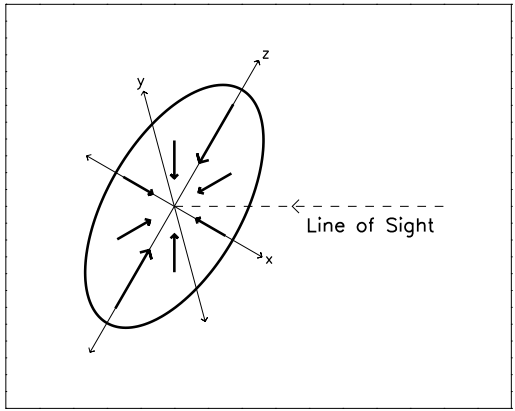}
  \caption{Diagram represents an elongated collapsing cluster which is viewed along a line of sight not coincident with a primary axis.  During the initial (collapse-like) phase of a cluster with subvirial initial conditions, the upper portion of the cluster will appear red-shifted while the lower portion of the cluster appears more blue-shifted.  This asymmetry results in a radial velocity gradient along the extent of the cluster.}
	\label{fig:LOSDiagram}
  \end{figure}

  Elongated subvirial clusters display a gradient in the radial velocity along the direction of elongation due to this initial collapse. For example, consider a cluster with its major axis in the $\hat{z}$ axis direction and let the system be observed along a line of sight that is less than $90^{\circ}$ from the major axis. For clarity, we define North and South to be the projected positive and negative $\hat{z}$ axes, respectively.  During the first half of a crossing time, the northern part of the cluster appears to be red-shifted away from the observer while the southern hemisphere is blue-shifted toward the observer.  The collapse of an elongated subvirial cluster naturally results in a North-South gradient in the radial velocities along the length of the cluster (see Fig. \ref{fig:LOSDiagram}). The magnitude and direction of the radial velocity gradient depends on the line of sight chosen to `observe' the simulated cluster (see $\S$\ref{sec:KinSign}).
  
  This kinematic signature in the radial velocities is present only if an elongated cluster is viewed along a line of sight that is \emph{not} coincident with a principal axis \emph{and} if the cluster has subvirial velocities.  When viewed along a minor axis, a subvirial elongated cluster displays no North-South gradient in the radial velocities, as the stars are preferentially moving perpendicular to the line of sight.  Viewed along the major axis, the portion of the cluster moving away from the observer is projected directly onto the region of the cluster moving toward the observer, and hence a gradient is not observed.  
  
  Likewise, a virial cluster does not display a strong radial velocity gradient, as it is not globally collapsing and the stellar velocities have no spatial correlation. A very slight gradient is observed in the virial clusters at early times; however, it is approximately $10$ times weaker than that observed in the subvirial clusters and is associated with the slight contraction of the simulated virial clusters as the stellar velocities are redistributed from the random isotropic distribution to a true virial distribution (see Fig. \ref{fig:DispvsTime}). 
  
\begin{figure}
  \epsscale{1.0}
  \plotone{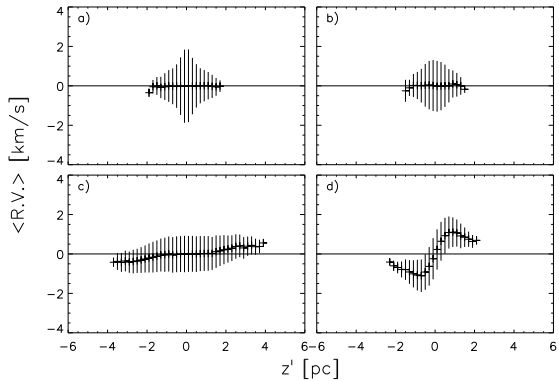}
  \caption{Radial velocity as a function of $\hat{z}'$ position for a differing geometries, initial velocity distributions, and projections for simulated embedded clusters.  The mean radial velocity (averaged over $0.2$ pc bins) is in indicated by the data points. The error bars correspond to the standard deviation of the radial velocity distribution within the bins.  The panels correspond to differing initial geometries, initial velocity distribution, and projections of simulated embedded clusters: (a) spherical cluster, with subvirial initial conditions, (b) elongated cluster, with subvirial initial conditions, and viewing angle of $0^{\circ}$ from major axis, (c) elongated cluster, with virial initial conditions, and viewing angle of $30^{\circ}$ from major axis, and (d) elongated cluster, with subvirial initial conditions, and viewing angle of $30^{\circ}$ from major axis.  The first distribution is shown at time $t = 1.5$ Myr after the start, whereas the other three cases are shown at $t = 2.1$ Myr.}
  \label{fig:Figure1-2}
  \end{figure}

  A cluster that is undergoing global expansion can also produce a gradient in the radial velocities similar to that observed in the collapsing elongated cluster (Figure \ref{fig:Figure1-2}, panel(d)). However, the signature will differ in shape because the stars that populate outermost spatial bins will not have bound orbits.  The stars will not be at the furthest point in their orbits and therefore will not have reduced velocities as they change direction and return to the cluster's center.  Therefore, instead of having an S-shaped kinematic signature, an expanding cluster will have a relatively flat gradient, with no turnover at the end points.  We also note that rotation of a virial cluster may produce a kinematic signature that is similar to the one observed in subvirial clusters, but only if the rotation takes place around an axis perpendicular to the line of sight.
    
    Figure \ref{fig:Figure1-2} illustrates the radial velocity signature for various projections of spherical and elongated subvirial and virial clusters.  In each panel the data points correspond to the mean radial velocity of the stars in $0.2$ pc bins and the length of the error bars correspond to the standard deviation of the velocities within each bin. For completeness, we note that this radial velocity gradient calculated from the simulated clusters is not sensitive to the choice of bin size for bin sizes ranging from $0.05$ to $1.0$ pc.  In individual clusters (such as the ONC), the lower limit of the bin size is determined by small number statistics.
    
    It is clear from this figure that the radial velocity signature is created by a combination of two effects: projection of an elongated cluster and subvirial initial velocities (the latter implies global collapse).  The bottom right panel shows the only cluster that demonstrates this kinematic signature.  This cluster is elongated, is not viewed along a primary axis, and is seeded with subvirial initial velocities.
    
    One way to characterize the strength of a radial velocity gradient is by using the amplitude of the cumulative velocity distribution, \acrv.  The cumulative radial velocity distribution is created by sorting the individual radial velocity measurements by the $\hat{z}'$ coordinate and then producing the cumulative distribution from this sorted data set.  We normalize this measure by both the number of stars in the cluster, $N$, and the velocity dispersion of the cluster, $\sigma_{RV}$.  This procedure results in a (dimensionless) normalized cumulative radial velocity distribution that can be more meaningfully compared across clusters of different sizes and velocity dispersions.  An additional advantage of the cumulative velocity distribution is that it is created from individual stellar radial velocities and thus is insensitive to the choice of bin size. 
    
    A cluster with no radial velocity gradient will have (on average) the same number of stars with positive and negative radial velocities with respect to the cluster's center of mass or mean radial velocity.  Therefore, the cumulative velocity distribution in such a cluster will fluctuate around zero and be relatively flat as a function of $z'$.
        
    On the other hand, a cluster with a strong gradient along $\hat{z}'$ will have preferentially more blue-shifted stars at the negative $\hat{z}'$ end of the cluster center and more red-shifted stars at the positive $\hat{z}'$ end of the cluster.  Therefore, the cumulative radial velocity distribution will be a decreasing function for $z' < 0$ and an increasing function for $z' > 0$, resulting in a dip in the cumulative radial velocity distribution.
    
\begin{figure}
  \epsscale{1.0}
  \plotone{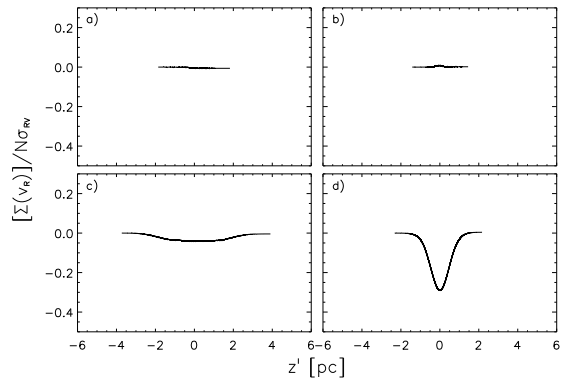}
  \caption{Cumulative radial velocity distribution (normalized by cluster size and velocity dispersion) summed along the $\hat{z}'$ axis of the projected cluster.  The panels correspond to differing initial geometries, initial velocity distribution, and projections of simulated embedded clusters: (a) spherical cluster, with subvirial initial conditions, (b) elongated cluster, with subvirial initial conditions, and viewing angle of $0^{\circ}$ from major axis, (c) elongated cluster, with virial initial conditions, and viewing angle of $30^{\circ}$ from major axis, and (d) elongated cluster, with subvirial initial conditions, and viewing angle of $30^{\circ}$ from major axis.  The first distribution is shown at time $t = 1.5$ Myr after the start, whereas the other three cases are shown at $t = 2.1$ Myr. }
	\label{fig:SumRV}
  \end{figure}

    The normalized cumulative radial velocity distributions for the clusters depicted in Figure \ref{fig:Figure1-2} are displayed in Figure \ref{fig:SumRV}.  The ranges on the $\hat{y}$ axes are held constant to emphasize how this distribution varies for subvirial and virial initial conditions.  Although the virial elongated cluster in panel (c) of Figure \ref{fig:Figure1-2} appears to have a slight gradient in the radial velocity, the cumulative distribution indicates that the strength of this signature is more than $7$ times weaker than in the subvirial elongated cluster.  The cumulative distribution for the virial cluster peaks at \acrvs $\sim 0.04$, whereas that of the subvirial cluster peaks at \acrvs $\sim 0.29$.  Note that the minimum is not observed in the subvirial spherical cluster, panel (a), which indicates that subvirial velocities alone are insufficient to produce this kinematic signature.  As a result, in general, both subvirial and non-spherical initial conditions are required to observe the kinematic signature.
    
  The key feature in the cumulative radial velocity distribution signature is neither the growth nor the decay of the function, but rather the peak produced by both a region of growth and then a region of decay.  While either the growth, or the decay, of the distribution can be mimicked by constructing the distribution in a velocity frame that is significantly different from the average velocity of the cluster, both cannot be created in the same frame.  For instance, if the cumulative radial velocity distribution for a virial cluster is constructed by shifting into a velocity frame that is significantly higher than the cluster's average velocity, then almost all stars contribute negative values to the distribution, and thus the distribution decreases as a function of $z'$.  For the purpose of this study, we construct all cumulative radial velocity distributions by shifting each stellar velocity by the median radial velocity.  
  
  We also note that a flat radial velocity gradient, such as that produced by a rotating virial cluster, or an expanding elongated cluster, can produce a peak in the cumulative radial velocity distribution similar to the one observed in the collapsing elongated cluster.  Therefore, both the kinematic signatures presented in Figure \ref{fig:Figure1-2} and the integrated distributions presented in Figure \ref{fig:SumRV} should be considered when looking for signs of collapse in an observed cluster's radial velocity structure.
    
  \subsection{Velocity Dispersions}
  
  Observations of kinematic signatures in young clusters can be compared to these simulated signatures only if the observed clusters are large enough that the binned radial velocity data do not suffer significantly from small number statistics.  We investigate trends in the total, radial, and tangential velocity dispersions in order to allow for comparison to smaller and less studied clusters for which only estimates for the total velocity dispersions are available.
  
  In both spherical and elongated subvirial clusters, the velocity dispersions increase as a function of time during the initial collapse phase and then decrease as the cluster evolves toward virial equilibrium.  Clusters that experience multiple collapse and re-expansion cycles before gas removal also display corresponding cycles in their velocity dispersions. The dispersions peak before the re-expansion phase and bottom out before the collapse phase, with the peak values becoming smaller with each successive cycle.  Virial clusters have velocity dispersions that increase only slightly during the first $1$ Myr, the star formation epoch, and then remain relatively flat through the rest of the embedded stage. These trends are similar to the ones depicted in Figure \ref{fig:DispvsTime}.  While an elongated cluster's total velocity dispersion does not depend on the projection, its radial and tangential velocity dispersions do.  This dependence on projection angle, $\theta$ is discussed in \S \ref{sec:ProjectionAngle}.

	\subsection{Kinematic Signatures}\label{sec:KinSign}
	
	As discussed in \S \ref{sec:RVStructure}, only clusters with subvirial initial velocities \emph{and} elongated geometries produce radial velocity gradients.  The kinematic signature evolves over time.  During the first $1$ Myr, the radial velocity gradient arises as stars become free to move through the gravitational potential of the cluster.  The gradient becomes larger during the first `free fall time' while the stars fall toward the cluster center.  As a majority of the stars pass through the cluster center, the radial velocity gradient decreases and eventually changes sign as the cluster re-expands.   Clusters with sufficiently small crossing times will undergo several gradient sign changes before gas removal at $t = 5$ Myr as the cluster size oscillates around its equilibrium size.  In addition, as the cluster evolves toward equilibrium, the strength of the kinematic signature decays.
			
	At $t = 5$ Myr, the embedding gas potential is removed from the cluster which then continues to evolve due to gravitational interactions between the stars. Depending on the specific kinematic status of the cluster at the time of gas removal, the signature considered here may or may not remain in the cluster.  If the gas is removed while a significant portion of the stars are in the re-expanding phase, the kinematic signature is amplified and remains in the cluster even after 5 Myr of evolution.  It is important to note that the details of the gas removal process may significantly affect the characteristics of the signature, and hence we refrain from making any strong conclusions about the kinematic signature's presence after gas removal.
		
		\subsubsection{Effect of Initial Virial Ratio}
		
				We compare centrally concentrated elongated clusters ($\rho \sim m^{-1}$) clusters with virial parameters $Q = 0.04$ and $0.15$ to determine how the departure from virial equilibrium effects the strength of the kinematic signature.   The clusters have $N = 2000$ members, a SFE of $50\%$, and axis parameters $a = b = 2$ pc and $c = 4$ pc.  We find that as the initial virial ratio (and the average starting stellar velocity) decreases, the strength of the kinematic signature increases.  Therefore, the larger the departure from virial equilibrium corresponds to a larger kinematic signature.   We measure the strength of the signature by the amplitude of the normalized cumulative radial velocity distribution \acrv, and find that the $Q = 0.04$ cluster had a maximum at \acrvs $\sim 0.33$ whereas the $Q = 0.15$ cluster had a maximum at \acrvs $\sim 0.20$.
																
				In addition, even though the $Q = 0.15$ cluster had larger initial velocities, the $Q = 0.04$ cluster has a larger average velocity dispersion over the embedded stage.  Specifically, the average velocity dispersions were $0.51$ km s$^{-1}$ and $0.47$ km s$^{-1}$ for the $Q = 0.04$ and $0.15$ clusters respectively.  
				
		\subsubsection{Effect of Projection Angle}\label{sec:ProjectionAngle}	
		
		For a subvirial elongated cluster, the strength of the radial velocity signature varies most strongly with projection angle.  This trend is due to two competing effects.  First, the most significant collapse occurs along the major axis of the cluster, the $\hat{z}$ axis.  The component of the velocity that is observed along the line of sight varies as the cosine of the projection angle $\theta$ between the $\hat{z}$ axis and the line of sight.   Therefore, the smaller the projection angle, the stronger the signature. Secondly, the radial velocity signature is only observed when there are significantly more red-shifted stars in a declination bin than blue-shifted stars (or vice versa).  Therefore, a larger projection angle $\theta$ will cause the red- and blue-shifted populations to be more spatially separated, whereas a small projection angle will result in a projected cluster whose red-shifted and blue-shifted populations appear to overlap.   
						
\begin{figure}
  \epsscale{0.85}
  \plotone{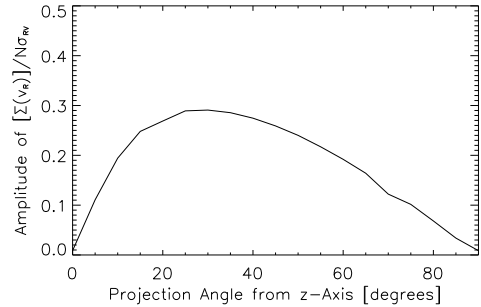}
  \caption{Amplitude of the normalized cumulative radial velocity distribution \acrv, as a function of projection angle $\theta$.  The data are shown for an elongated cluster with axis ratios $(1:1:2)$, uniform stellar and gas densities and subvirial initial velocities at the time $t = 2.1$ Myr.}
  \label{fig:ProAngle}
  \end{figure}

		Figure \ref{fig:ProAngle} displays the amplitude of the normalized cumulative radial velocity distribution \acrv, as a function of projection angle $\theta$, for the uniform density elongated cluster with $a = b = 2$ pc and $c = 4$ pc at $t = 2.1$ Myr.  A projection angle between $25$ and $35$ degrees from the $\hat{z}$ axis produces the strongest radial velocity signature.  The peak of the distribution is relatively broad, however, with a full width at half maximum of approximately $50$ degrees.  
		
		During collapse, the dispersion in the radial velocities $\sigma_{RV}$ is a decreasing function of the projection angle $\theta$ from the $\hat{z}$ axis, whereas the velocity dispersion along the $\hat{z}'$ axis $\sigma_{\parallel, z'}$ is an increasing function of $\theta$.    This result occurs because the dispersions are geometrically related to each other by $\sigma_{RV}(\theta) = \sigma_{\parallel, z'}(90-\theta)$.   These trends are representative of the fact that during the collapse phase, the velocity dispersions are largest along the principal axis. 
		
\begin{figure}
  \epsscale{0.9}
  \plotone{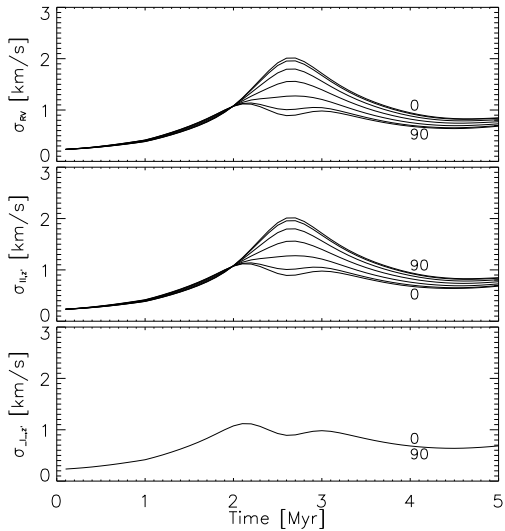}
	\caption{The evolution of the velocity dispersions in an elongated subvirial cluster during the embedded stage for various projection angles.   The top panel displays the radial velocity dispersion $\sigma_{RV}$  as a function of time for $\theta = 0, 15, 30, 45, 60, 75,$ and $90$ degrees.  The middle panel displays the velocity dispersion parallel to the $\hat{z}'$ axis, $\sigma_{\parallel, z'}$ for the same projection angles.  Note that $\sigma_{RV} (\theta) = \sigma_{\parallel, z'} (90-\theta)$ by definition.  The bottom panel shows the velocity dispersion in the plane-of-sky perpendicular to the $\hat{z}'$ axis, $\sigma_{\perp, z'}$.  On account of the $x,y$ symmetry of a prolate spheroid, $\sigma_{\perp, z'}$ does not vary as a function of $\theta$. The data presented here are from a uniform spheroid cluster with subvirial initial velocities and axis parameters $a = b = 2$ pc and $c = 4$ pc.}
  \label{fig:VelocityDispersions}
  \end{figure}

		In contrast, the plane-of-sky velocity dispersion perpendicular to the $\hat{z}'$ axis does not depend on the projection angle.  This result is due to the $x,y$ symmetry of the prolate spheroid.  Figure \ref{fig:VelocityDispersions} displays $\sigma_{RV}$ (top panel), $\sigma_{\parallel, z'}$ (middle panel), and $\sigma_{\perp, z'}$ (bottom panel) for the same cluster shown in Figure \ref{fig:ProAngle}.   Each curve corresponds to a different choice of projection angle, $\theta = 0, 15, 30, 45, 60, 75,$ and $90$ degrees.  The relationship between $\sigma_{RV}$ and $\sigma_{\parallel, z'}$ described above is apparent in Figure \ref{fig:VelocityDispersions}.
		
    \subsubsection{Effect of Cluster Elongation}
		
		To observe differences caused by the amount of elongation, we compare two subvirial clusters with differing axis ratios.  Specifically, we compare a cluster with an aspect ratio of $0.5$ ($a = b = 2$ pc, $c = 4$ pc) to one with an aspect ratio of $0.25$ ($a = b = 1$ pc, $c = 4$ pc).  Each of the clusters has 2700 members,  a density profile $\rho \sim m^{-1}$, a star formation efficiency of $33\%$, and a virial parameter $Q = 0.04$. 
		
		We find that the more elongated cluster evolves on a slightly shorter time scale and that the mean radial velocities in a given declination bin are larger.  As a result, the kinematic signature discussed in $\S$\ref{sec:RVStructure} is stronger.  For comparison, the maximum mean radial velocity in a declination bin is approximately $1$ km s$^{-1}$ for the less elongated cluster and $2$ km s$^{-1}$ for the more strongly elongated cluster.  Therefore the overall change in radial velocity is approximately $4$ km s$^{-1}$ over the length of the cluster in the strongly elongated geometry.  The average velocity dispersion is also larger in the more elongated cluster, $\sigma = 1.60$ km s$^{-1}$, compared to the less elongated cluster where $\sigma = 1.18$ km s$^{-1}$.

		\subsubsection{Effect of Initial Density Distribution}
		
		The kinematic signatures also vary as a function of initial density distribution.  We investigate this variation by comparing subvirial ($Q = 0.04$) elongated cluster with uniform density distributions to those with centrally concentrated $\rho \sim m^{-1}$ density profiles.  Each cluster contains $N = 2000$ members, has an SFE of $50\%$, and has axis parameters $a = b = 2$ pc and $c = 4$ pc.
		
		Elongated clusters that are more centrally concentrated have central regions which evolve on shorter time scales than the outer regions.  This trend can be seen by considering the gravitational contraction of an centrally concentrated sphere with radius $r_0$ and density profile $\rho \sim r^{-1}$ .   The collapse is inside-out because free fall time for a particle initially at a position $r$ is given by $t_{ff} \sim \rho^{-1/2} \sim \sqrt{r}$, and so the central regions of the system evolve on shorter timescales.  \cite{Goodman1983MNRAS} showed that this general argument holds true for spheroidal clusters as well.   
		
		On account of this inside-out collapse, the center of a subvirial cluster will collapse and begin to re-expand while the outskirts of the cluster are still collapsing.  As a result, the red and blue shifted populations will not be as spatially separated as in the uniform density cluster shown in Figure \ref{fig:Figure1-2}, Panel (d).  Instead, the generally red-shifted $z'> 0$ portion of the cluster will contain many blue-shifted stars near the center, and this addition acts to blue-shift the average radial velocity in the central $z'$ bins.  On the other side of the cluster's center, red-shifting may occur for the same reason, resulting in a kinematic signature that is flat or even oppositely sloped (with respect to the general trend) near $z' = 0$.
		
		In addition, the inside-out collapse causes the average velocity dispersion during the embedded stage to be higher in centrally concentrated clusters than in uniform density clusters.  This effect occurs because the peak in the velocity dispersion, which occurs during the cluster's collapse and re-expansion, is broader (in time) in the $\rho \sim m^{-1}$ clusters than in the $\rho \sim \rho_0$ clusters.  In a centrally concentrated spheroid, the stellar members reach their highest speeds at a time that depends on their initial position within the cluster.  In contrast, in a uniform spheroid, the stellar members reach their highest speeds at roughly the same time.  Therefore, the velocity dispersion in a centrally concentrated cluster remains higher for a longer period of time.    This finding has observational consequences: initially centrally concentrated clusters are more likely to be seen with high velocity dispersions.       
				
		\subsubsection{Effect of Star Formation Efficiency}
		
		To investigate the effect of SFE on the kinematic signatures observed, we compare elongated clusters with SFE of  $17\%, 33\%,$ and $50\%$. Each cluster has a $\rho \sim m^{-1}$ stellar density profile, axis parameters $a = b = 2$ pc and $c = 4$ pc, and subvirial $Q = 0.04$ initial velocities.   The clusters with SFEs of $17\%$ and $13\%$ have $N = 2700$ stars, whereas the cluster with a $50\%$ SFE has $N = 2000$ stars. 
		
		Clusters with lower star formation efficiencies evolve on shorter timescales due to the higher velocities associated with the deeper potential wells of the gas component.  In addition, both the amplitude \acrvs and the total velocity dispersion increase as SFE decreases.
				
		\subsubsection{Effect of Extinction due to High Column Density}
		
		Most very young clusters are still embedded in the molecular cloud from which they formed.  In cases where optical spectra are used to determine stellar kinematics, extinction from the embedding clouds can potentially change the observed kinematic signatures in these clusters.  For example, consider a cluster that is roughly shaped like a prolate spheroid with its North end tipped toward the observer. The Southern region is farther away, and is collapsing toward the center of mass, and thus appears blue-shifted with respect to the stars in the northern region. If even a modest fraction of the cluster members that are farthest from the observer are not included in the observed data set due to extinction by the embedding cloud, this blue-shifted population would appear less blue-shifted than if it were not obscured.  These observed selection effects may be mitigated by using H or K band spectra to select targets.  

		As an example, consider an embedding cloud that has a dust opacity of $\kappa_V = 200$ cm g$^{-1}$ at optical wavelengths and a mean molecular mass $\mu = 2.4 m_H$.  In order to have an optical depth with the value of unity, the minimum column density is required to be $N_{col}\approx 1.28 \times 10^{21}$ cm$^{-2}$.  A typical molecular cloud has number densities ranging from $n = 10^2 - 10^3$ cm$^{-3}$.  Therefore, if the sources are being observed through more than $0.4 - 4$ pc of molecular cloud, they would be undetected and hence removed from the sample.
		
\begin{figure}
  \epsscale{1.0}
  \plotone{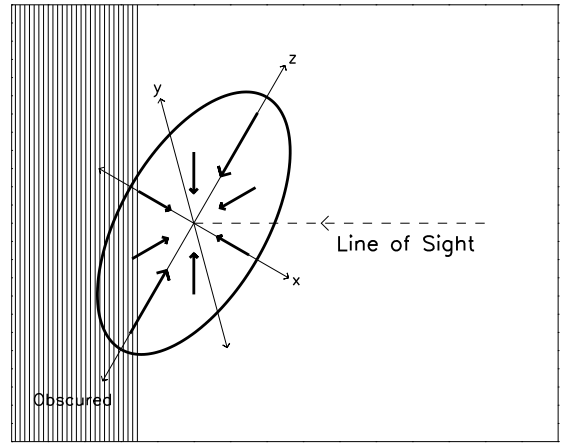}
	\caption{Same as Figure \ref{fig:LOSDiagram}, but indicates portion of cluster that is assumed to be obscured due to extinction.}
	\label{fig:ObscPlane}
  \end{figure}

\begin{figure}
  \epsscale{0.7}
  \plotone{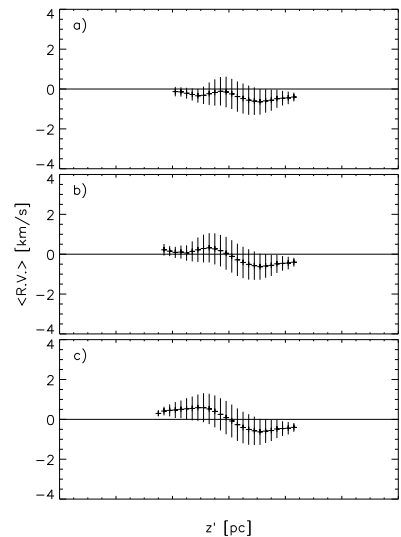}
	\caption{Each panel displays the radial velocity as a function of $\hat{z}'$ position in a uniform density subvirial elongated cluster for different locations of the obscuring plane.  The mean radial velocity (averaged over $0.2$ pc bins) is in indicated by the data points. The error bars correspond to the standard deviation of the radial velocity distribution within the bins. In the top panel, $d_{obs} = 0.6$ pc and $32\%$ of the cluster members are removed from the sample.  The radial velocity signature in the middle panel is observed when $d_{obs} = 1.4$ pc and $12\%$ of the cluster members are thus obscured.  The bottom panel is the radial velocity signature produced when all stars are included in the sample. Even modest amounts of extinction can wash out the structure of the kinematic signature.}
	\label{fig:Extinction}
  \end{figure}

	  To test this hypothesis, we re-analyzed the simulated cluster data by omitting stars that are behind an ``obscuring plane'' (see Fig. \ref{fig:ObscPlane}) which includes a column density threshold.  In this model, the obscuring plane is introduced with an orientation normal to the line of sight and at a distance $0 \leq d_{obs} \leq 5$ pc beyond the center of the cluster from the observer.  We found that if only a modest number of stars in the Southern region of the cluster are removed from the sample (due to the fact that they are beyond the obscuration plane) and their velocities are not included in the velocity versus position plot, the Southern portion of the cluster will have a significantly less steep velocity gradient compared to the Northern region.  Specifically, if $10 - 15\%$ of the stellar population is unobservable, the kinematic signature in the southern half of the simulated cluster is completely washed out.  Figure \ref{fig:Extinction} displays the radial velocity signatures for a cluster with various amounts of the stars extincted.  In the top panel, $d_{obs} = 0.6$ pc and $32\%$ of the cluster members are removed from the sample.  The radial velocity signature in the middle panel is observed when $d_{obs} = 1.4$ pc and $12\%$ of the cluster members are thus obscured.  For comparison, the bottom panel is the radial velocity signature produced when all stars are included in the sample.
	  
  \section{Comparison to Observations} \label{sec:CompareObs}
  
  Recent kinematic studies of the Orion star-forming region have determined radial velocities for a large sample ($\sim 1200$) of visible sources within the region.  In this section we discuss the kinematic results of these studies in light of the numerical simulations presented above.  
  
  The ONC region is a good environment in which to look for kinematic signatures such as the ones observed in our simulated clusters for many reasons.  First, the ONC has a large population and is close enough to be well studied so that many of the cluster members have measured radial velocities.  The large sample size of the cluster allows the data to be binned in declination while maintaining a reasonable number of data points in each bin, so that the results suffer only mildly from small number statistics.  Furthermore, the ONC is visibly elongated in projection \citep{Hillenbrand1998ApJ}, which is one of the requirements for the kinematic signature to be observed.    
  
  Finally, although opinions vary, there are significant arguments supporting the assertion that the ONC region as a whole is young and is estimated to be less than one crossing time old.  Observations of the ONC by \cite{Furesz2008ApJ} and more recently by T09 identify spatially coherent kinematic structure in the stellar distribution that closely matches the observed kinematic structure of the gas in the region as measured in $^{13}$CO by \cite{Bally1987ApJ}. The authors argue that this correlation indicates the cluster is not dynamically relaxed and that in fact, the region is less than a crossing time old. The results of the previous section indicate that the strength of the kinematic signature peaks before the cluster is a crossing time old.  Therefore, the strongest observed signature will occur in clusters that are less than a crossing time old, such as the ONC.  
   
	  \subsection{Observations}

	  The radial velocities cited in this study were determined from multiple observations using Hectochelle on MMT and MIKE fibers at Magellan. The Hectochelle observations from \cite{Furesz2008ApJ} have been combined with additional Hectochelle and MIKE observations (T09) to produce a list of $\sim 1200$ sources with radial velocity measurements in the region surrounding the ONC.  A detailed discussion of the observations, data reduction, spectral fitting, and radial velocity measurements is provided in the observational papers.  In addition, much care has been taken to identify binaries and non-cluster members from the ONC data, and this procedure is detailed in T09.
	  
    \subsection{Kinematic Signatures}

\begin{figure}
  \epsscale{1.0}
  \plotone{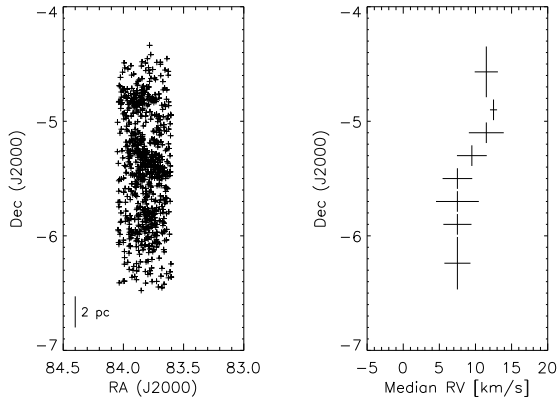}
	\caption{Left panel displays the positions of stars in the ONC with radial velocities measured by \citep[][T09]{Furesz2008ApJ} that are spatially associated with the filament of dense gas.  In the right panel, the median radial velocity of the stellar population is plotted for each declination bin.  Bin width is indicated by the vertical extent of the bars.  The horizontal error bars indicate the velocity dispersion in each bin, where the dispersion is defined by the distribution's full width at half maximum divided by $2\sqrt{2 \ln 2}$.  There is a noticeable radial velocity gradient in the Northern ONC, which is qualitatively similar to the one observed in the simulated subvirial clusters. }
	\label{fig:VZAvg_Orion}
  \end{figure}

  As discussed in these observational papers \citep[][T09]{Furesz2008ApJ}, the stars and gas in the region surrounding the ONC show similar North-South velocity gradients.  We consider the ONC members that are near the molecular cloud filament by selecting the RA range 84.0 to 83.5, and remove identified binaries from the sample.   The distribution of sources is shown in the left panel of Figure \ref{fig:VZAvg_Orion}.  In the right panel of Figure \ref{fig:VZAvg_Orion} the binned median velocities are plotted as a function declination.  To calculate the median velocity in each declination bin, a histogram of the radial velocities is created. The median is calculated ignoring all bins that have less than half of the maximum value in the distribution.   The medians calculated without localizing on the filament differ from those shown here by a few 0.5 km/s shifts to the red or blue.
  
  Figure \ref{fig:VZAvg_Orion} is analogous to Figure \ref{fig:Figure1-2} for the simulated cluster data.  The North-South velocity gradient is clear, though the gradient is steeper in the northern part of the ONC (north of declination $\sim -5.5$) than it is in the southern region.  The observed kinematic structure in the ONC region is qualitatively similar to the kinematic structure of elongated subvirial clusters viewed off axis (bottom right panel of Fig. \ref{fig:Figure1-2}. 
  
\begin{figure}
  \epsscale{1.0}
  \plotone{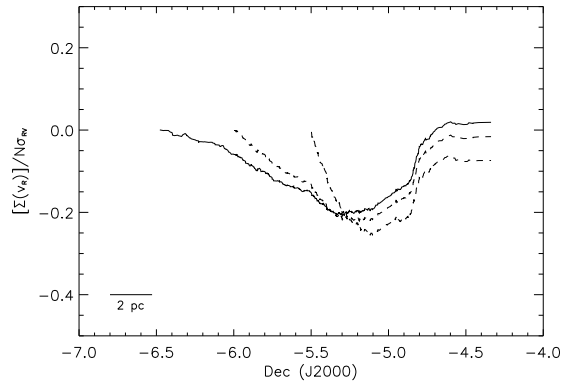}
	\caption{Normalized cumulative radial velocity distribution summed along declination in the Orion star forming region.  This cumulative distribution is created from the ONC members with radial velocities within 3 $\sigma$ of the median radial velocity of the distribution.  From \cite{Furesz2008ApJ}, the radial velocity dispersion in this region of the ONC is $\sigma = 3.1$ km s$^{-1}$.  The cumulative radial velocity distributions for the stars north of declination $-6$ and $-5.5$ are indicated by the dashed lines.}
	\label{fig:SumRV_Orion}
  \end{figure}

  Figure \ref{fig:SumRV_Orion} presents the cumulative radial velocity distribution observed in the ONC (analogous to Figure \ref{fig:SumRV}) for the entire sample (solid curve) as well as the distribution for only those stars north of declination $-6$ and $-5.5$ (dashed curves). This distribution is composed of the radial velocities of the ONC members that have velocities within 3 $\sigma$ of the median radial velocity of the distribution of the median velocity, where $\sigma = 3.1$ km s$^{-1}$ \citep{Furesz2008ApJ}.  The constrained data set (without the declination cuts) includes approximately $89\%$ of the sources shown in the left panel of Figure \ref{fig:VZAvg_Orion}.  To produce these plots, the radial velocities were shifted by the median observed velocity and normalized by the velocity dispersion of and number of stars in the data set.  This shift and normalization allows the observed data to be directly compared to the cumulative radial velocity distribution in the simulated clusters, where the velocities are measured in the center of mass reference frame.  The kinematic signature peak observed in our simulated clusters is also observed in the ONC, over the extent of the cluster, as well as over just the northern portion.  The ONC data is significantly less smooth than the theoretical results.  This difference is mostly due to the summing of 50 simulations used to produce the theoretical results shown in Figure \ref{fig:SumRV}.
  
  
	  \subsection{Comparison to Simulation Results}
	  
	  The kinematic structure observed in the ONC may be understood in light of the simulated cluster kinematics.  The correlation between the motion of the stars and the gas in the ONC indicates that the region is fairly young and that the stars and gas are still collapsing.  The observed radial velocity gradient also supports the theory that the ONC is dynamically young and that the region is less than a crossing time old.  Also, the signature suggests the stars in the region are formed with subvirial velocities, an initial condition required by the simulations to produce the structure observed.  The ONC region is elongated and the combination of elongation, subvirial initial velocities, and extremely young dynamical age can account for the kinematic structure observed as demonstrated by the simulations presented in this study.  This interpretation requires that the Northern region of the ONC be closer to the observer than the Southern region.  However, at this time, the distance to the ONC as a whole is still poorly constrained \citep[][and references therein]{Jeffries2007MNRAS} and determining distance differences on the order of $10$ pc between the Northern and Southern regions of the ONC is not yet possible.
	  	  
  The kinematic signatures observed in the ONC are qualitatively similar to those observed in the elongated subvirial clusters.  The simulated cluster that is physically most similar to the ONC region is the centrally concentrated elongated cluster with axis ratio: $2:2:8$ pc and a low star-formation efficiency of $\sim 17 \%$.  This cluster displays a kinematic signature similar in magnitude to the observed signature in the ONC.  Specifically, at $t =0.7$ Myr, the simulated cluster has a radial velocity gradient of $\sim 2$ km s$^{-1}$ pc$^{-1}$, and \acrvs $\sim 0.15$.  In addition, the simulated cluster has a radial velocity dispersion of $\sim 3$ km s$^{-1}$. In comparison, the Orion data have a radial velocity gradient of $\sim 5$ km s$^{-1}$ over $0.6^{\circ}$ in declination; at a distance of $420$ pc, the parameters are approximately a gradient of $1.1$ km s$^{-1}$ pc$^{-1}$, \acrvs $\sim 0.2$ and a radial velocity dispersion of $3.1$ km s$^{-1}$.
	  
	  Proper motion studies in Orion (such as the one conducted by \citealt{Jones1988AJ}) found plane-of-sky velocity dispersions along the cluster's major axis that are somewhat larger than the dispersions perpendicular to the major axis.  In the notation adopted in this paper, $\sigma_{\parallel, z'} = 2.63 \pm 0.9$ km s$^{-1}$ and $\sigma_{\perp, z'} = 2.03 \pm 0.11$ km s$^{-1}$.  \cite{Furesz2008ApJ} determined a radial velocity dispersion of $\sigma_{RV} = 3.1$ km s$^{-1}$.  This relatively large difference between $\sigma_{RV}$ and $\sigma_{\perp, z'}$ is indicative of some type of global collapse because virial elongated clusters have at most modest differences between $\sigma_{RV}$ and $\sigma_{\perp, z'}$.  In our simulated virial clusters $1 < \sigma_{RV}/\sigma_{\perp, z'} \leq 1.3$, whereas in the subvirial clusters $1 < \sigma_{RV}/\sigma_{\perp, z'} \leq 2.3$.
	  
  The ONC region displays significant mass segregation among the most massive stars ($m \gtrsim 2 \msun$) which reside in the Trapezium at the center of the ONC.  Previous numerical studies of the ONC cluster indicate that the mass segregation must be primordial as the cluster is not old enough for dynamical mass segregation to account for the presence of the Trapezium \citep{Bonnell1998MNRAS}.  Our simulated clusters begin with somewhat different initial conditions, by focusing on subvirial initial velocities and cluster elongation, but arrive at the same conclusion:  the region is not old enough for dynamical mass segregation to have taken place.  In the simulated clusters, we compared the radii that enclosed $25, 50,$ and $75\%$ of the stellar mass in different mass bins.  The results of these simulations showed no difference in the radii of different mass bins during the first $1-3$ Myr of evolution, indicating that dynamical mass segregation had not yet occurred.  Subvirial initial velocities will result in a higher central density at earlier times on account of global collapse, but even with an increased density and thus higher interaction rate, the ONC is still too young for significant mass segregation to have occurred. 

 In addition to the lack of mass segregation in the spatial distribution of stars within the cluster, no significant differences in the kinematic distributions are observed as a function of stellar mass in the simulations.  The masses of stars in the Orion region are not well enough constrained to compare this result of the simulations to the observations at this time.  However, future studies of young stars in this region may yield mass information that, combined with kinematic data, will be able to test this prediction.
	  
	  As seen in Figure \ref{fig:VZAvg_Orion}, the radial velocity gradient apparent in the northern region is not as evident in the region south of the ONC.  The asymmetry may indicate that the center of the gravitational collapse is not the Trapezium, but rather is located slightly north of the Trapezium. The collection of cumulative radial velocity distributions displayed in Figure \ref{fig:SumRV_Orion}, however, indicate that the signature is present even if the center of the gravitational collapse is assumed to be somewhat north of the Trapezium.  The asymmetry could also be due to a north-south asymmetry in the initial conditions that is not well represented by the idealized simulations of \S \ref{sec:SimResults}.  The star-forming region north of the ONC is located at the edge of the integral-shaped molecular cloud, whereas the region south of the ONC extends into a larger complex of molecular clouds.  Therefore the dynamics in the northern region is better suited to our approximate of an isolated elongated centrally concentrated cluster than those of the south.

\section{Conclusion} \label{sec:Conclusion}

This study presents $N$-body simulations that explore the kinematic signatures produced by asphericity and subvirial initial velocities in young embedded clusters.  We have identified a robust kinematic signature, in particular a gradient in the radial velocities, which is naturally produced by elongated subvirial clusters.  We characterize the properties of this kinematic signature as a function of initial conditions.  Specifically, we compare the kinematics of evolving spherical and aspherical clusters and with both virial and subvirial initial conditions.  We consider changes in the kinematic signature due to differing amounts of cluster elongation, star-formation efficiency and departure from virial equilibrium, and discuss possible sample bias due to extinction in observed clusters.  Finally we compare the signature displayed in our simulated clusters to kinematic data from the ONC.  The main results of our work are summarized as follows:

\begin{itemize}
\item Elongated clusters with subvirial initial velocities display a gradient in the radial velocities as a function of projected position along the cluster.  Both aspherical initial geometries and subvirial initial velocities are required to produce this kinematic signature (see Fig. \ref{fig:Figure1-2}).  
\item The strength of the kinematic signature increases during the first free-fall time as the cluster collapses, and then decreases as the stars pass through the center of the cluster.  The gradient changes sign as the cluster re-expands, and the amplitude of the cumulative radial velocity distribution decreases (see Fig. \ref{fig:SumRV}) as the cluster evolves (see also Fig. \ref{fig:DispvsTime}).
\item The strength of the kinematic signature varies most sensitively as a function of the projection angle, as measured from the major axis of the cluster.  The signature is weakest at projection angles of $0^{\circ}$ or $90^{\circ}$ and strongest at projection angles of $\sim 25 - 35^{\circ}$ (see Fig. \ref{fig:ProAngle}). 
\item Making the cluster more elongated increases the gradient of the radial velocity across the cluster while leaving the velocity dispersions roughly similar.  
\item The initial stellar (and gas) density distribution affects the rate at which the cluster evolves and thus the time scale on which the kinematic signature evolves.  In addition, centrally concentrated clusters have higher velocity dispersions than uniform density clusters.  
\item The kinematic signature is sensitive observational selection effects.  Extinction in an embedded cluster can preferentially deselect stars further from the observer and will thus affect the kinematic signature.  A modest amount of extinction that removes $10\%$ of the cluster members will wash out the kinematic signature in the region of extinction (see Fig. \ref{fig:ObscPlane}).
\item The asymmetric kinematic signature is qualitatively similar to the observed kinematics of the Orion star-forming region, suggesting that in addition to being elongated (as observed) a significant fraction of the cluster members started with subvirial initial velocities (see \S \ref{sec:CompareObs}).  The large gradient observed ($\sim 1$ km s$^{-1}$ pc$^{-1}$) indicates that the cluster is dynamically young.  This result is consistent with previous independent claims \citep[e.g.,][]{Bonnell1998MNRAS}.\end{itemize}

Previous studies have shown that the star-formation efficiency of clusters and the initial velocity distribution have significant effects on the long term evolution of a cluster \citep{Adams2000ApJ, Boily2003MNRAS,Adams2006ApJ}, i.e., the cluster bound fraction and stellar interaction rates.  This work emphasizes the fact that these conditions affect the short term evolution of the cluster as well.  In addition, this work is significantly different from previous studies in that it considers non-spherical initial conditions in both the stellar and gaseous components in young embedded clusters.  Although many examples of clusters with non-spherical geometries exist and a considerable amount of work has considered orbits of individual stars within axisymmetric and triaxial potentials, little work has been done theoretically to understand how these clusters, taken as a whole, differ from spherical clusters.    We have shown that elongated clusters naturally produce observable kinematic signatures that depend on the initial starting velocities of the cluster members and the projection of the cluster onto the plane of the sky.  These signatures may allow us to identify, or at least constrain, the initial conditions in star-forming clusters that still retain most of their embedding gas.

The kinematic signatures found in the elongated subvirial cluster simulations may shed some light on kinematics within nearby young clusters. As one example, this work shows that the gradient produced in simulated clusters is qualitatively similar to that observed in the radial velocities of the ONC members.  Although the ONC kinematic data display some additional asymmetry not observed in the simulated clusters, this asymmetry is likely to be due to the more complicated environment and feedback processes which are not well represented by the simulation.   Nevertheless, the general structure and magnitude of the observed kinematic signature in the ONC may be explained theoretically, provided that the stars are formed with subvirial initial velocities.  

Current instruments such as the Hectochelle and MIKE have helped make large spectroscopic studies of many objects more efficient and have produced many radial velocity studies of stellar clusters as well as larger globular clusters and dwarf galaxies.  These instruments have provided insight into the kinematics of astronomical objects ranging from the rotation of individual stars to the large scale dynamics of galaxies.  In the next decade, the GAIA mission will provide the opportunity to study the three dimensional kinematics of clusters with exquisite detail.  For example, proper motions and radial velocities will be measured for stars in ONC brighter than 20th magnitude. This study will result in accurate transverse velocities to combine with the radial velocities supplemented by ground-based observations.  As more detailed kinematic data become available for the ONC and other young clusters, they should be used in conjunction with cluster simulations such as those presented here to understand the initial conditions necessary to produce the observed kinematics.  A more detailed understanding of the initial conditions can then inform cluster formation and evolution models, with the overarching goal of better understanding star and planet formation within young stellar clusters.

We thank G. F{\H u}r{\'e}sz, T. Megeath, and P. Myers for useful suggestions and discussions over the course of this study.  We also thank the referee, Dr. Nick Moeckel, for his constructive comments which strengthened the presentation of our results.  Support for F.A. and E.P. was provided by the University of Michigan through the Michigan Center for Theoretical Physics, and by the Spitzer Space Telescope Theoretical Research Program (1290776).  L.H. and J.T. acknowledge support provided by a NASA grant (NNX08AI39G) and by the University of Michigan.





\end{document}

%% file: tab1.tex
  \begin{deluxetable*}{lrllc}
	\tablewidth{0pt}
  \tabletypesize{\footnotesize}
  \tablecaption{Stellar Distributions\label{tab:StarDist}}
  \tablehead{
  \colhead{Description} & \multicolumn{3}{c}{Density Profile} & \colhead{Parameters} } 
  \startdata
  Spherical & $\rho(\xi)$ & $= \left\{ \begin{array}{l}
    \rho_0/\xi, \\ 0, \end{array} \right. $ & $\begin{array}{l} 0 < \xi \leq 1 \\ \xi > 1 \end{array} $ 
    & $\xi = r/r_0$ \\
  Uniform Spheroid & $\rho(m)$ & $= \left\{ \begin{array}{l}
    \rho_0, \\ 0, \end{array} \right. $ &  $\begin{array}{l} 0 \leq m \leq 1 \\ m > 1 \end{array} $ 
    & $ \begin{array}{c} m^2 = (x/a)^2 + (y/b)^2 + (z/c)^2 \\ a = b < c \end{array}$ \\
  $1/m$ Spheroid & $\rho(m)$ & $= \left\{ \begin{array}{l} 
    \rho_0/m, \\ 0, \end{array} \right. $ & $\begin{array}{l} 0 < m \leq 1\\ m > 1 \end{array} $
    & $ \begin{array}{c} m^2 = (x/a)^2 + (y/b)^2 + (z/c)^2 \\ a = b < c \end{array}$ \\
  \enddata
  \end{deluxetable*}

%% file: tab2.tex
  \begin{deluxetable*}{lrllc}
  \tablewidth{0pt}
  \tabletypesize{\footnotesize}
  \tablecaption{Embedding Gas Distributions\label{tab:GasDist}}
  \tablehead{
  \colhead{Description} & \multicolumn{3}{c}{Density Profile} & \colhead{Parameters} } 
  \startdata
  Spherical & $\rho(\xi)$ & $= \begin{array}{l} \rho_0/\xi(1+\xi)^3, \end{array}$ & $\begin{array}{l}0 < \xi \end{array}$ 
    & $\xi = r/r_0$ \\
  Uniform Spheroid & $\rho(m)$ & $= \left\{ \begin{array}{l}
    \rho_0, \\ 0, \end{array} \right.$ & $\begin{array}{l} 0 \leq m \leq 2 \\ m > 2 \end{array}$
    & $\begin{array}{c} m^2 = (x/a)^2 + (y/b)^2 + (z/c)^2 \\ a = b < c \end{array}$ \\
  $1/m$ Spheroid & $\rho(m)$ & $= \begin{array}{l} \rho_0/m,\end{array}$ & $ \begin{array}{l} 0 < m \end{array}$
    & $ \begin{array}{c} m^2 = (x/a)^2 + (y/b)^2 + (z/c)^2 \\ a = b < c \end{array}$ \\
  \enddata
  \end{deluxetable*}

%% file: ms.bbl
\begin{thebibliography}{24}
\expandafter\ifx\csname natexlab\endcsname\relax\def\natexlab#1{#1}\fi

\bibitem[{{Aarseth}(2001)}]{Aarseth2001NewAstro}
{Aarseth}, S.~J. 2001, New Astron., 6, 277

\bibitem[{{Adams}(2000)}]{Adams2000ApJ}
{Adams}, F.~C. 2000, \apj, 542, 964

\bibitem[{{Adams} {et~al.}(2007){Adams}, {Bloch}, {Butler}, {Druce}, \&
  {Ketchum}}]{Adams2007ApJ}
{Adams}, F.~C., {Bloch}, A.~M., {Butler}, S.~C., {Druce}, J.~M., \& {Ketchum},
  J.~A. 2007, \apj, 670, 1027

\bibitem[{{Adams} \& {Fatuzzo}(1996)}]{Adams1996ApJ}
{Adams}, F.~C. \& {Fatuzzo}, M. 1996, \apj, 464, 256

\bibitem[{{Adams} {et~al.}(2006){Adams}, {Proszkow}, {Fatuzzo}, \&
  {Myers}}]{Adams2006ApJ}
{Adams}, F.~C., {Proszkow}, E.~M., {Fatuzzo}, M., \& {Myers}, P.~C. 2006, \apj,
  641, 504

\bibitem[{{Allen} {et~al.}(2007){Allen}, {Megeath}, {Gutermuth}, {Myers},
  {Wolk}, {Adams}, {Muzerolle}, {Young}, \& {Pipher}}]{Allen2007PPV}
{Allen}, L., {Megeath}, S.~T., {Gutermuth}, R., {Myers}, P.~C., {Wolk}, S.,
  {Adams}, F.~C., {Muzerolle}, J., {Young}, E., \& {Pipher}, J.~L. 2007,
  Protostars and Planets V, 361

\bibitem[{{Andr{\'e}}(2002)}]{Andre2002ApSS}
{Andr{\'e}}, P. 2002, \apss, 281, 51

\bibitem[{{Bally} {et~al.}(1987){Bally}, {Stark}, {Wilson}, \&
  {Langer}}]{Bally1987ApJ}
{Bally}, J., {Stark}, A.~A., {Wilson}, R.~W., \& {Langer}, W.~D. 1987, \apjl,
  312, L45

\bibitem[{{Binney} \& {Tremaine}(1987)}]{BT1987Book}
{Binney}, J. \& {Tremaine}, S. 1987, {Galactic Dynamics} (Princeton, NJ:
  Princeton University Press)

\bibitem[{{Boily} \& {Kroupa}(2003)}]{Boily2003MNRAS}
{Boily}, C.~M. \& {Kroupa}, P. 2003, \mnras, 338, 673

\bibitem[{{Bonnell} \& {Davies}(1998)}]{Bonnell1998MNRAS}
{Bonnell}, I.~A. \& {Davies}, M.~B. 1998, \mnras, 295, 691

\bibitem[{{F{\H u}r{\'e}sz} {et~al.}(2008){F{\H u}r{\'e}sz}, {Hartmann},
  {Megeath}, {Szentgyorgyi}, \& {Hamden}}]{Furesz2008ApJ}
{F{\H u}r{\'e}sz}, G., {Hartmann}, L.~W., {Megeath}, S.~T., {Szentgyorgyi},
  A.~H., \& {Hamden}, E.~T. 2008, \apj, 676, 1109

\bibitem[{{Goodman} \& {Binney}(1983)}]{Goodman1983MNRAS}
{Goodman}, J. \& {Binney}, J. 1983, \mnras, 203, 265

\bibitem[{{Hernquist}(1990)}]{Hernquist1990ApJ}
{Hernquist}, L. 1990, \apj, 356, 359

\bibitem[{{Hillenbrand} \& {Hartmann}(1998)}]{Hillenbrand1998ApJ}
{Hillenbrand}, L.~A. \& {Hartmann}, L.~W. 1998, \apj, 492, 540

\bibitem[{{Jeffries}(2007)}]{Jeffries2007MNRAS}
{Jeffries}, R.~D. 2007, \mnras, 376, 1109

\bibitem[{{Jones} \& {Walker}(1988)}]{Jones1988AJ}
{Jones}, B.~F. \& {Walker}, M.~F. 1988, \aj, 95, 1755

\bibitem[{{Kirk} {et~al.}(2007){Kirk}, {Johnstone}, \& {Tafalla}}]{Kirk2007ApJ}
{Kirk}, H., {Johnstone}, D., \& {Tafalla}, M. 2007, \apj, 668, 1042

\bibitem[{{Lada} \& {Lada}(2003)}]{Lada2003ARAA}
{Lada}, C.~J. \& {Lada}, E.~A. 2003, \araa, 41, 57

\bibitem[{{Miller} \& {Scalo}(1979)}]{Miller1979ApJS}
{Miller}, G.~E. \& {Scalo}, J.~M. 1979, \apjs, 41, 513

\bibitem[{{Myers} {et~al.}(1991){Myers}, {Fuller}, {Goodman}, \&
  {Benson}}]{Myers1991ApJ}
{Myers}, P.~C., {Fuller}, G.~A., {Goodman}, A.~A., \& {Benson}, P.~J. 1991,
  \apj, 376, 561

\bibitem[{{Peretto} {et~al.}(2006){Peretto}, {Andr{\'e}}, \&
  {Belloche}}]{Peretto2006AA}
{Peretto}, N., {Andr{\'e}}, P., \& {Belloche}, A. 2006, \aap, 445, 979

\bibitem[{{Porras} {et~al.}(2003){Porras}, {Christopher}, {Allen}, {Di
  Francesco}, {Megeath}, \& {Myers}}]{Porras2003AJ}
{Porras}, A., {Christopher}, M., {Allen}, L., {Di Francesco}, J., {Megeath},
  S.~T., \& {Myers}, P.~C. 2003, \aj, 126, 1916

\bibitem[{{Ryden}(1996)}]{Ryden1996ApJ}
{Ryden}, B.~S. 1996, \apj, 471, 822

\bibitem[{{Walsh} {et~al.}(2004){Walsh}, {Myers}, \& {Burton}}]{Walsh2004ApJ}
{Walsh}, A.~J., {Myers}, P.~C., \& {Burton}, M.~G. 2004, \apj, 614, 194

\end{thebibliography}
